\documentclass{research4cacm}
\usepackage{epsfig,color}

\usepackage{lmodern} 
\usepackage{listings}
\newcommand{\code}[1]{{\bf\texttt{#1}}}

\newtheorem{theorem}{Theorem}[section]

\newtheorem{lemma}[theorem]{Lemma}

\begin{document}


\title{Egel --- Graph Rewriting with a Twist}
\numberofauthors{1}
\author{
\alignauthor
M.C.A. (Marco) Devillers\\
       \email{marco(dot)devillers(at)gmail.com}
}
\maketitle


\begin{abstract}
Egel is an untyped eager combinator toy language. Its primary purpose 
is to showcase an abstract graph-rewriting semantics allowing a
robust memory-safe construction in C++.
Though graph rewriters are normally implemented by elaborate
machines, this can mostly be avoided with a change in the representation of 
term graphs. With an informal inductive argument, that 
representation is shown to always form directed acyclic graphs.
Moreover, this graph semantics can trivially be extended to allow
exception handling and cheap concurrency.
Egel, the interpreter, exploits this semantics with a straight-forward
mapping from combinators to reference-counted C++ objects.
\end{abstract}


\section{Introduction}

It all started with Lisp. Except that it didn't. 
Throughout history, people have been interested in mechanizing math and,
more recently, mathematical approaches to programming. Countless
researchers have contributed to this ideal, most are forgotten, but
certain influential milestones can be identified which tell a
story from symbolic evaluation to graph-driven combinatorial
rewriting.

Lisp\cite{mccarthy:lisp1} put the representation and symbolic 
evaluation of expressions first and coupled that with a versatile 
operational semantics; 
the language and ideas behind it remain influential to this day.
The first work on the mechanical evaluation of non-strict languages was
laid down by Landin\cite{landin:secd} resulting research which put
closures first.
Turner's work on SASL\cite{turner:sasl} diverged from that and
concentrated on SK-combinator-driven evaluation culminating in
the typed and lazily reduced Miranda\cite{turner:miranda}.
Combinator-driven lazy graph rewriting spurred a number of 
abstract machines such as the Spineless Tagless Graph Machine\cite{spj:stgm}
behind GHC/Haskell and the Parallel ABC Machine\cite{plasmeijer:clean} for Clean.

But while graph rewriting is a pleasingly elegant means to give an
operational semantics to a term-rewriting language, ultimately it
was deemed too slow and compilers for functional languages now
usually invest a great deal into compiling to more traditional
schemes.
However, because graph rewriting is such a simple model with
some exceptional properties, it allows for trivialized implementations
of term-rewriting languages.

Egel exploits a novel view on eager graph rewriting to implement
a term-rewriting language in a robust and memory-safe manner in
C++, at the cost of performance.

\section{Graph Rewriting}

\begin{figure}[h]
\begin{center}
\epsfbox{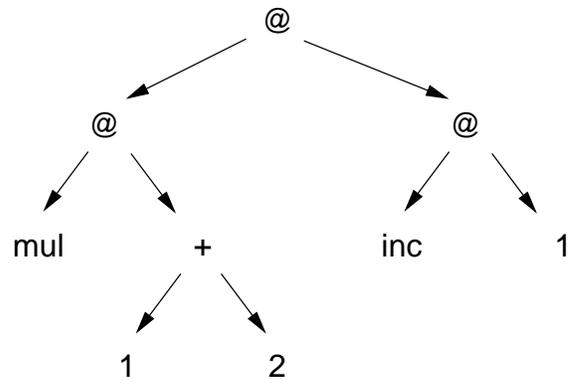}
\caption{standard graphical representation}
\label{figure:termA}
\end{center}
\end{figure}

The notion of graph rewriting starts with the observation that usually
a term of a language can be given a pictorial representation.
In figure~\ref{figure:termA}, the traditional tree representation of
the term \code{mul (1 + 2) (inc 1)}, a running example, is given. 
The \code{@} node depicts application.

\begin{figure}[h]
\begin{center}
\epsfbox{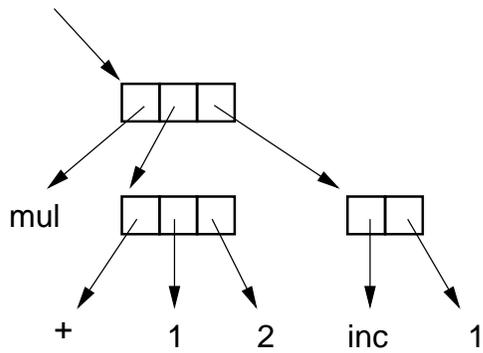}
\caption{thunked representation}
\label{figure:termB}
\end{center}
\end{figure}

However, the representation of a term graph in computer memory is often
slightly different. In figure~\ref{figure:termB}, a standard `thunked'
representation of \code{mul (1 + 2) (inc 1)}, a thunk is an array
of pointers to constants and combinators.
Note that the \code{@} application node is gone conforming to
that storing unnecessary application nodes would be too 
costly regarding both storage and performance.

Given the thunked representation of term graphs a straight-forward
approach towards an evaluator would be to introduce primitives
and graph-manipulation code for combinators combined with a stack machine 
which holds redexes to rewrite, traces of that can be found in 
both the G-Machine and the PABC machine.

\begin{figure}[h]
\begin{center}
\epsfbox{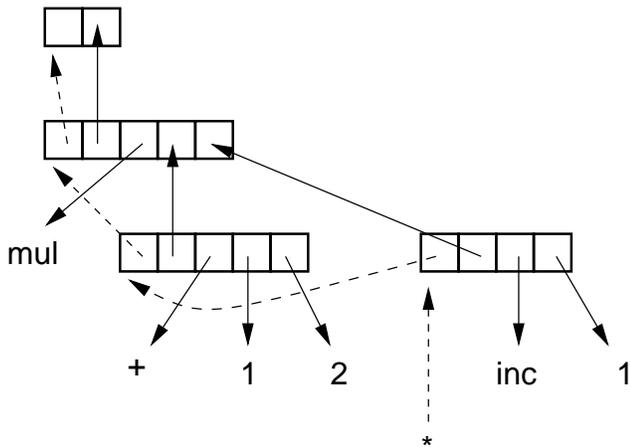}
\caption{twisted representation}
\label{figure:termC}
\end{center}
\end{figure}

Instead of that, Egel terms are compiled to a twisted representation,
as shown in figure~\ref{figure:termC}, bypassing the need
for a stack. Thunks are extended at the front with two pointers,
one pointer points to what to do --rewrite-- next and another pointer
where to store the result.
The \code{*} root node points to what will be rewritten first.

The cost of allocating extra pointers in a thunk may seem wasteful
but shouldn't be more burdensome than allocating a thunk and some
stack space. Though the performance benefits of this approach are 
completely undone by Egel's wasteful idiomatic C++ implementation.

The chain of redexes to rewrite makes explicit that the reduction
order is strict or eager. Arguments to functions are rewritten first,
in right-to-left order, after which the function is applied.
Reduction is performed by repeatedly rewriting the top root pointer.

\begin{figure}[h]
\begin{center}
\epsfbox{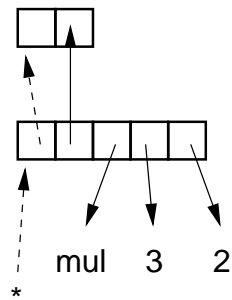}
\caption{after evaluation of two arguments}
\label{figure:termD}
\end{center}
\end{figure}

Figure~\ref{figure:termD} shows the term after the first two
arguments of \code{mul} are rewritten. Note that reduced
arguments are always pointed towards; i.e., should form a tree.

\begin{figure}[h]
\begin{center}
\epsfbox{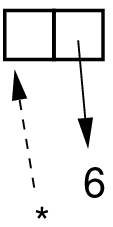}
\caption{final term}
\label{figure:termE}
\end{center}
\end{figure}

The fully reduced term is shown in figure~\ref{figure:termE}. At
this point the runtime can be called by the root rewriting pointer
and might, for instance, print the result.

The clue of this paper: Where the first picture originally starts of
with a tree, or with sharing a directed acyclic graph (DAG), each
other figure still is a tree, or DAG; changing representation or 
rewriting kept this invariant.

\section{The Egel Language}

The Egel language is an experimental front-end with the previously
described graph semantics. It's not thoroughly discussed here,
with two examples just enough of a taste of the language is given
to understand a follow-up informal argument.
Below, an example Egel script implementing the Fibonacci function.

\begin{verbatim}
import "prelude.eg"

namespace Fibonacci (
    using System
  
    def fib =
        [ 0 -> 1
        | 1 -> 1
        | N -> fib (N - 2) + fib (N - 1) ]
)

using Fibonacci

def main = fib 5
\end{verbatim}

Superficially, Egel isn't much different from other functional
programming languages. As shown, scripts can include other 
scripts, it has namespaces, and definitions of functions may
be recursive.

What is slightly different is that functions are defined
with guarded lambda abstractions which are directly mapped
to (unnamed) combinators by the interpreter.

The example script below shows how lists are defined and used.

\begin{verbatim}
namespace List (
    data nil, cons

    def ++ =
        [ nil YY -> YY
        | (cons X XX) YY -> cons X (XX ++ YY) ]
)
\end{verbatim}

Noteworthy is that a good approximation of Egel is to think
of it as a lambda calculus with constants where constants
may compose; e.g., \code{(1 2)} is a legal term in Egel.

That feature is exploited to introduce the notion of lists;
in the example script two constants \code{nil} and \code{cons}
are defined which, as any other constant, may be applied 
to any number of arguments. Guarded abstractions are then 
used to define (recursive) functions which may decompose their
arguments.

\section{Directed Acyclic Graph Property}

This section discusses the heart of this paper, an informal
argument.

\begin{theorem}
Egel terms in the runtime always form a tree, or directed
acyclic graph.
\end{theorem}

The proof is an inductive argument which relies on a
property of the front-end Egel language. From now on,
tree is written where also directed acyclic graph is meant.

\begin{lemma}
Fully reduced expressions always form a tree.
\end{lemma}

This is fundamentally a property of the front-end
language since combinators could, in principle, rewrite
terms in the runtime to anything. However, it is assumed
that combinators are the result of the translation of 
code in the front end, i.e., complex expressions of 
guarded anonymous abstractions. Since abstractions can
only take apart and reassemble complex trees, with
some confidence this property holds. 

\begin{lemma}
The chain of redexes always forms a tree, even during rewriting.
\end{lemma}

This is an inductive argument. First, the base step, the intial term
populated with the \code{main} combinator forms a tree, 
which is trivially true through inspection.

Then, the inductive step, if the chain of redexes forms a tree,
then rewriting won't change that. This holds because a rewrite
can result in either of two things: Either the fully
reduced result (a tree) is placed in a receiving thunk and
rewriting proceeds with the next redex, and that is trivially
again a tree. Or, the chain of redexes is expanded with a new number 
of redexes, conforming to the translation of the right-hand-side of
a guarded abstraction, and that must form a tree.

The inductive step was checked by comparing some source code to
their byte code translation.

\section{C++}

The Egel interpreter has a runtime which is a mapping of the above
graph rewrite machinery to idiomatic C++ code. Input
scripts are translated in a very trivial manner where,
after lambda-lifting, all guarded abstractions are mapped
to combinators, and each combinator is mapped to a C++
object. Combinators, or C++ objects, can contain byte
code comprised of simple graph manipulation instructions.
The runtime consists of nothing more than a
collection of C++ objects which rewrite each other;
each thunk pointed to by the root pointer is 
simply called in a trampoline loop.

One major benefit is that this appoach is `provably'
memory safe since combinators/objects are natively
reference-counted objects allocated with the advised
Resource Acquisition Is Initialization (RAII) scheme; i.e.,
\code{malloc} and \code{new} are 
completely avoided in the Egel interpreter source
code. A major drawback is
performance, idiomatic C++ incurs a hefty cost in
allocation and number of indirections. In short,
this approach is robust but slow.

This simple semantics does allow for
more experimentation. Concurrency is trivially
implemented by inserting nodes into the graph
which start rewriting in parallel to each other.
Moreover, the scheme of arrows
to redexes to rewrite also allows for a
trivial implementation of exceptions and 
exception handling: each thunk is extended
again with a pointer to the expression which
holds an exception handler thunk to be applied
when an exception occurs.

Localized reasoning is seen as a major benefit
of this model; i.e., there's no stack and there's
no need for a complex garbage collection scheme
including, for instance, a global mark and collect
phase. The hope is that that will map better to 
heavily concurrent architectures in the future
than other approaches.

\section{Conclusions}

Egel is a toy language which serves as a front-end
to novel graph-rewrite machinery. This graph-rewrite
machinery is able to express computation as the 
sole result of cooperating nodes in a directed 
acyclic graph.

Egel is implemented in C++ where combinators
are directly mapped to objects in a memory-safe
manner.

Localized reasoning, no stack, no global mark and
collect, is hoped to map well to future massively
concurrent microprocessor architectures.

Because this operational model is that simple, it
allows for a lot of experimentation, which will
be documented in other notes.

\bibliographystyle{abbrv}
\bibliography{egelbib}
\balancecolumns
\end{document}